\documentclass[aps,prd,nofootinbib,twocolumn,amssymb,eqsecnum,notitlepage]{revtex4-1}

\usepackage{bm}
\usepackage{amsmath,amsthm,amssymb}
\usepackage{color}
\usepackage[dvipdfmx]{graphicx}
\usepackage{enumitem}
\usepackage{hyperref}
\usepackage{array}
\usepackage[english]{babel}
\usepackage{tensor}
\usepackage{comment}
\usepackage[normalem]{ulem}
\usepackage[dvipsnames]{xcolor}
\allowdisplaybreaks

\def\be{\begin{equation}}
\def\ee{\end{equation}}
\def\ba{\begin{eqnarray}}
\def\ea{\end{eqnarray}}

\def\tg{\tilde{g}}
\def\ta{\tilde{a}}
\def\tY{\tilde{Y}}
\def\tb{\tilde{b}}

\begin{document}

\title{Tracker and scaling solutions in DHOST theories}

\author{Noemi Frusciante$^1$, Ryotaro Kase$^2$, Kazuya Koyama$^3$,
Shinji Tsujikawa$^2$ and Daniele Vernieri$^4$}

\affiliation{\smallskip
$^1$Instituto de Astrof\'isica e Ci\^encias do Espa\c{c}o, 
Faculdade de Ci\^encias da Universidade de Lisboa,  Campo Grande, PT1749-016 Lisboa, Portugal
\smallskip\\
$^2$Department of Physics, Faculty of Science, Tokyo University of Science,
1-3, Kagurazaka, Shinjuku-ku, Tokyo 162-8601, Japan
\smallskip\\
$^3$Institute of Cosmology \& Gravitation, University of Portsmouth, Dennis Sciama Building, Portsmouth, PO1 3FX, United Kingdom
\smallskip\\
$^4$Centro de Astrof\'isica e Gravita\c c\~ao  - CENTRA,
Departamento de F\'isica, Instituto Superior T\'ecnico - IST,
Universidade de Lisboa - UL, Av. Rovisco Pais 1, 1049-001 Lisboa, 
Portugal}

\begin{abstract}

In quadratic-order degenerate higher-order scalar-tensor (DHOST) theories compatible with 
gravitational-wave constraints, we derive the most general Lagrangian 
allowing for tracker solutions characterized by $\dot{\phi}/H^p={\rm constant}$, 
where $\dot{\phi}$ is the time derivative of a scalar field $\phi$, $H$ is 
the Hubble expansion rate, and $p$ is a constant. 
While the tracker is present up to the cubic-order Horndeski Lagrangian 
$L=c_2X-c_3X^{(p-1)/(2p)} \square \phi$, where 
$c_2, c_3$ are constants and $X$ is
the kinetic energy of $\phi$, the DHOST interaction breaks this structure for $p \neq 1$.
Even in the latter case, however, there exists an approximate tracker solution 
in the early cosmological epoch with the nearly constant field equation of state 
$w_{\phi}=-1-2p\dot{H}/(3H^2)$.
The scaling solution, which corresponds to $p=1$, is the unique case in which 
all the terms in the field density $\rho_{\phi}$ and the pressure 
$P_{\phi}$ obey the scaling relation $\rho_{\phi} \propto P_{\phi} \propto H^2$. 
Extending the analysis to the coupled DHOST theories with the field-dependent 
coupling $Q(\phi)$ between the scalar field and matter, we show that the scaling 
solution exists for $Q(\phi)=1/(\mu_1 \phi+\mu_2)$, 
where $\mu_1$ and $\mu_2$ are constants. 
For the constant $Q$, i.e., $\mu_1=0$, we derive fixed points of the dynamical system 
by using the general Lagrangian with scaling solutions. 
This result can be applied to the model construction of late-time cosmic acceleration 
preceded by the scaling $\phi$-matter-dominated epoch.

\end{abstract}

\pacs{98.80.-k,98.80.Jk}

\maketitle

\section{Introduction}

There have been numerous attempts to modify or extend General Relativity (GR) 
at large distances \cite{review1,review2,review3,review4,review5,review6}. 
One of such motivations is to explain the observational 
evidence of late-time cosmic acceleration by introducing a new ingredient 
beyond the scheme of standard model of particle physics. 
The simple candidate for such a new degree of freedom (DOF) is 
a scalar field $\phi$ \cite{quin1,quin2,quin3,quin4,quin5,quin6,quin7}, 
which has been widely exploited to describe the 
dynamics of dark energy.

The theories in which the scalar field is directly coupled to gravity 
(with two tensor polarized DOFs)  are generally 
called scalar-tensor theories \cite{Brans,Fujii}. 
It is known that Horndeski theories \cite{Horndeski} are 
the most general scalar-tensor theories with second-order equations of motion \cite{Def11,KYY,Char11}. 
The second-order property ensures the absence of an Ostrogradsky 
instability \cite{Ostrogradsky} 
associated with a linear dependence of the Hamiltonian arising from extra DOFs. 

Horndeski theories can be extended to more general theoretical schemes without 
increasing the number of propagating DOFs \cite{Zuma}.
For example, Gleyzes-Langlois-Piazza-Vernizzi (GLPV) expressed the 
Horndeski Lagrangian in terms of scalar quantities arising in 
the 3+1 decomposition of spacetime \cite{building}
and derived a beyond-Horndeski Lagrangian without imposing 
two conditions Horndeski theories obey \cite{GLPV}. 
The Hamiltonian analysis in the unitary gauge showed that 
the GLPV theories do not increase the number of 
DOFs relative to those in Horndeski gravity \cite{Hami1,Hami2,Hami3}.

One can further perform a healthy extension of Horndeski theories by keeping  
one scalar and two tensor DOFs. 
Even if Euler-Lagrange equations contain derivatives higher than 
second order in the scalar field and the metric, it is possible to maintain the same number of 
 propagating DOFs by imposing the so-called degeneracy conditions of their 
Lagrangians \cite{Langlois1,Langlois2,CKT,Ben2, Ben}. 
They are dubbed degenerate higher-order scalar-tensor (DHOST) 
theories, which encompass GLPV theories as a special case. 
The absence of an extra DOF was confirmed by the Hamiltonian 
analysis \cite{Langlois2,Hami3} as well as by the field definition 
linking to Horndeski theories \cite{CKT, Ben2, Cris16}.

The DHOST theories contain the products of covariant derivatives of the field 
which are quadratic and cubic in $\nabla_{\mu} \nabla_{\nu} \phi$, say, 
$(\square \phi)^2$ and $(\square \phi)^3$, respectively. 
If we apply the DHOST theories to dark energy and adopt the bound of 
the speed $c_t$ of gravitational waves constrained from the GW170817 
event \cite{GW170817} together with the electromagnetic counterpart \cite{Goldstein}, 
the Lagrangians consistent with $c_t=1$ (in the unit where the speed of
light is equivalent to 1) are up to quadratic in $\nabla_{\mu} \nabla_{\nu} \phi$ 
with one of the terms vanishing ($A_1=0$) \cite{Matas} 
among six coefficients of derivative interactions. 
From the degeneracy conditions there are three constraints among the 
other five coefficients \cite{Cris17, LSYN, Cris18}, so we are left with two  
quadratic-order free functions. 
If we take into account the decay of 
gravitational waves to dark energy \cite{Cremi18}, 
we have an additional constraint on the quadratic-order 
Lagrangian\footnote{This assumes that the effective theories of dark energy 
are valid up to the energy scale corresponding to the gravitational-wave 
frequency observed by 
LIGO/Virgo ($f \sim 100$~Hz) \cite{deRham18}. }. 
Hence there is one free DHOST interaction containing the term
$B_4(\phi,X)R$ (where $R$ is the Ricci scalar) besides the 
Horndeski Lagrangian $L=G_2(\phi,X)-G_3(\phi,X) \square \phi$ 
up to cubic order. 

If we apply shift-symmetric Horndeski theories to dark energy, 
there are self-accelerating solutions preceded by 
a constant tracker equation of state $w_{\phi}$ 
with $\dot{\phi} \propto H^p$ ($p$ is a constant).
For example, the covariant Galileon \cite{Gali1,Gali2} 
gives rise to the value $w_{\phi}=-2$ with $p=-1$ 
during the matter era \cite{DT10,DT10b}, but it is disfavored from 
the joint data analysis of supernovae type Ia, 
cosmic microwave background, and baryon acoustic oscillations \cite{NDT}. 
The extended Galileon proposed in Ref.~\cite{DT11} can accommodate
the tracker equation of state $w_{\phi}$ closer to $-1$, 
in which case the model can be consistent with 
the observational data \cite{DT11b}. 
In DHOST theories, (approximate) tracker solutions were found
for particular models \cite{Cris18,CKLSN}, 
but the general conditions for its existence have been unknown.

In Horndeski theories, there is a special kind of tracker 
called the scaling solution \cite{Pedro,CLW,Liddle,Sahni,Skordis,Dodelson,Barreiro,Piazza,Tsuji04,Amendola06,Gomes1,Gomes2,Alb,ABDG} 
along which the field density $\rho_{\phi}$ is 
proportional to the background matter density $\rho_{m}$. 
If the scalar field has a constant coupling $Q$ with matter, 
the scaling solution 
satisfying the relation $\dot{\phi} \propto H$ exists for the 
cubic-order Horndeski Lagrangian $L=Xg_2(Y)-g_3(Y) \square \phi$, 
where $g_2, g_3$ are arbitrary functions of 
$Y=Xe^{\lambda \phi}$ ($\lambda$ is a constant) \cite{Frusciante:2018aew}.
In this case, it is possible to construct viable dark energy 
models with a scaling $\phi$-matter dominated epoch ($\phi$MDE). 
In DHOST theories, the conditions for realizing the 
scaling solution were not derived yet.

In this paper, we will derive the Lagrangian allowing for tracking 
and scaling solutions in DHOST theories compatible with observational 
constraints of gravitational waves. We impose the condition 
$\dot{\phi} \propto H^p$ by assuming that the quantity 
$h=-\dot{H}/H^2$ is nearly constant.
For $p \neq 1$, we show the existence of approximate tracker 
solutions characterized by the field equation of state  
$w_{\phi} \simeq -1+2ph/3$
in the early cosmological epoch. 
The scaling solution with the power $p=1$ is the special case in which 
the exact scaling behavior of the field density 
($\rho_{\phi} \propto \rho_m \propto H^2$) can be realized 
without assuming the dominance of $\rho_{m}$ over $\rho_{\phi}$.
We also extend the analysis to the case in which 
a field-dependent coupling $Q(\phi)$ between $\phi$ and matter 
is present and obtain the most general Lagrangian allowing for scaling solutions.

This paper is organized as follows. 
In Sec.~\ref{Sec:back}, we derive the background equations of motion in 
DHOST theories in the presence of the field-dependent coupling 
$Q(\phi)$ with matter. 
In Sec.~\ref{Sec:lag1}, we constrain the forms of DHOST Lagrangians 
allowing for the existence of tracker and scaling solutions for $Q=0$. 
In Sec.~\ref{Sec:lag2}, we obtain the most general Lagrangian 
with scaling solutions for the field-dependent 
coupling $Q(\phi)$ and also derive the fixed points of the dynamical system 
for constant $Q$. 
Sec.~\ref{consec} is devoted to conclusions.

\section{Background equations in DHOST theories}
\label{Sec:back}

Let us consider the quadratic-order DHOST theories 
given by the action~\cite{Langlois1,Langlois2,CKT,Ben2, Ben}: 
\be
{\cal S} = 
\int d^{4}x \sqrt{-g} \left( \frac{R}{2}+L \right)+
{\cal S}_m \left( \phi, g_{\mu \nu} \right)\,,
\label{action}
\ee
where $g$ is the determinant of metric tensor $g_{\mu \nu}$, 
$R$ is the Ricci scalar, and 
\be
L=G_2(\phi,X)-G_3(\phi,X) \square \phi+B_4(\phi,X)R
+A_4(\phi,X)Z.
\label{Lag}
\ee
Here, we use the unit where the reduced Planck mass 
$M_{\rm pl}$ is equivalent to 1.
The functions $G_2$, $G_3$, $B_4$, $A_4$ depend 
on $\phi$ and 
$X=-\nabla_{\mu}\phi\nabla^{\mu}\phi/2$, 
with the covariant derivative operator $\nabla_{\mu}$ and 
the d'Alembert operator $\square=\nabla^{\mu}\nabla_{\mu}$.
The quantity $Z$ is defined by 
\be
Z=
\nabla^{\mu}\phi \nabla_{\nu} \phi 
\nabla_{\mu}\nabla_{\rho}\phi
\nabla^{\rho} \nabla^{\nu}\phi\,.
\ee
The function $A_4$ is related to $B_4$ according to  
\ba
A_4=\frac{3B_{4,X}^2}{1+2B_4}\,,
\label{A4B4}
\ea
with the notation 
$B_{4,X} \equiv \partial B_4/\partial X$. The full DHOST theories contain 
the other four Lagrangians 
$L_1=A_1(\phi,X) \nabla_{\mu} \nabla_{\nu} \phi
\nabla^{\mu} \nabla^{\nu} \phi$, 
$L_2=A_2(\phi,X) \left( \square \phi \right)^2$, 
$L_3=A_3(\phi,X) \left( \square \phi \right)
\nabla^{\mu}\phi\nabla^{\nu}\phi \nabla_{\mu} \nabla_{\nu} \phi$, 
and $L_5=A_5(\phi,X)(\nabla^{\mu}\phi\nabla^{\nu}\phi 
\nabla_{\mu} \nabla_{\nu} \phi)^2$.
Requiring that the speed $c_t$ of gravitational waves is 
equivalent to 1, it follows that $A_1=0$ \cite{Matas}. 
The degeneracy conditions constrain
the coupling $A_2$ to be 0 and the functions $A_3$ and $A_5$ 
are related to each other according to $A_5=-2A_3(B_{4,X}+A_3X)/(1+2B_4)$.
To avoid the decay of gravitational waves to 
dark energy perturbations~\cite{Cremi18}, 
these functions are constrained to be $A_3=0=A_5$. 
On using the other degeneracy condition, we end up with the 
Lagrangian (\ref{Lag}) with the particular relation (\ref{A4B4}). 

The theory (\ref{Lag}) can be also obtained from the 
cubic-order Horndeski Lagrangian 
$L=P(\phi,X)+Q(\phi,X) \square \phi+f(\phi)R$ after performing 
an invertible conformal transformation 
$g_{\mu \nu} \to C(\phi,X) g_{\mu \nu}$~\cite{Cremi18}. 
We also note that the GLPV theories~\cite{GLPV} correspond to 
$A_3=-A_4=-B_{4,X}/X \neq 0$ and $A_5=0$, so they 
do not belong to the Lagrangian (\ref{Lag}). 
In other words, the GLPV theories do not satisfy the 
bound arising from the decay of gravitational waves 
to dark energy.

For the matter action ${\cal S}_m$, we consider a barotropic 
perfect fluid which can be coupled to the scalar field $\phi$.
In scalar-tensor theories without vector propagating DOFs, 
the matter sector can be described by the Schutz-Sorkin action~\cite{Sorkin,DGS,GPcosmo,GPGeff,DeFelice:2016ucp,Frusciante:2018aew}:
\be
{\cal S}_m=-\int d^4 x \left[ \sqrt{-g}\,\rho_{m} 
\left( n, \phi \right)+J^{\mu} \nabla_{\mu} \ell \right]\,,
\label{Sm}
\ee
where the matter density $\rho_m$ depends on the fluid 
number density $n$ as well as on $\phi$. 
The four vector $J^{\mu}$ is related to $n$, as 
$n=\sqrt{J^{\mu}J^{\nu} g_{\mu \nu}/g}$, and 
$\ell$ is a scalar quantity. 
We define the coupling between $\phi$ and matter, as 
\be
Q(\phi) \equiv \frac{\rho_{m,\phi}}{\rho_m}\,,
\ee
where $\rho_{m,\phi} \equiv \partial \rho_m/\partial \phi$.

To study the background cosmological dynamics, 
we consider a flat Friedmann-Lema\^{i}tre-Robertson-Walker 
spacetime given by the line element
\be
ds^2=-N^2(\hat{t})d\hat{t}^2+a^2(\hat{t}) 
\delta_{ij}dx^i dx^j\,,
\ee
where $N(\hat{t})$ is a lapse, and $a(\hat{t})$ is a scale factor. 
The Lagrangian in the action (\ref{action}) is given by 
\ba
L &=& G_2+ \left( \ddot{\phi}+3H\dot{\phi} 
\right) G_3+6\left( 2H^2+\dot{H} \right)B_4 
\nonumber \\
&&
-\frac{3B_{4,X}^2}{1+2B_4} 
\dot{\phi}^2 \ddot{\phi}^2,
\label{Lag2}
\ea
with $\sqrt{-g}=Na^3$ and $H \equiv \dot{a}/a$, where 
a dot represents the derivative with respect to 
$t \equiv \int N d \hat{t}$. 

For the matter sector, the temporal 
component of $J^{\mu}$ is related to the background 
number density $n_0$, as $J^0=n_0a^3$, so the matter 
action is expressed as
\be
{\cal S}_m=-\int d^4x\,a^3 \left[ N\rho_m(n_0, \phi)
+n_0 \frac{d\ell}{d\hat{t}} \right]\,.
\ee
The variations of ${\cal S}_m$ with respect to $n_0$ 
and $\ell$ lead to $\dot{\ell}=-\rho_{m,n}$ where 
$\rho_{m,n} \equiv \partial \rho_m/\partial n$, and 
\be
\dot{n}_0+3H n_0=0\,,
\label{n0eq}
\ee
respectively.

Varying the total action (\ref{action}) with respect to 
$N$ and $a$, we obtain the modified Friedmann equations:
\ba
& & 3H^2=\rho_{\phi}+\rho_m\,,\label{back1} \\
& & 2\dot{H}+3H^2=-P_{\phi}-P_m\,.\label{back2}
\ea
Here, $\rho_{\phi}$ and $P_{\phi}$ correspond to the 
field density and pressure defined, respectively, by 
\begin{widetext}
\ba
\hspace{-0.5cm}
\rho_{\phi}
&=&\dot{\phi}^2 G_{2,X}-G_2
-\dot{\phi}^2 \left( G_{3,\phi} 
-3H \dot{\phi} G_{3,X} \right)-6H^2 B_4
-6H \dot{\phi} \left( B_{4,\phi}+H \dot{\phi} B_{4,X} 
+\dot{\phi}^2 B_{4,X \phi} \right) \nonumber \\
& &+\frac{3}{\dot{\phi}} B_1 \left(1+2B_4 \right) 
\left[2\dddot{\phi} B_1+2H \ddot{\phi} \left( 3B_1 
-1 \right)-\frac{3}{\dot{\phi}} \ddot{\phi}^2 B_1 
\right]+6B_1 (1+2B_4) \left( \dot{H}+3H^2 \right)  \nonumber \\
& & +6 H \dot{\phi} \left[ 2B_1 B_{4,\phi}+B_{1,\phi} 
(1+2B_4) \right]+6B_1 \left[ B_1 \ddot{\phi}^2 B_{4,X}+
2B_1 \ddot{\phi} B_{4,\phi}
+(1+2B_4) \ddot{\phi} \left( \ddot{\phi} B_{1,X}
+2B_{1,\phi} \right) \right]\,,
\label{rhophi}\\
\hspace{-0.5cm}
P_{\phi}
&=& G_2-\dot{\phi}^2 \left( G_{3,\phi}+\ddot{\phi}\,G_{3,X} \right)
+2B_4 \left( 2\dot{H}+3H^2 \right)
+4H \dot{\phi}B_{4,\phi}+2 \ddot{\phi} \left( B_{4,\phi}
+2H \dot{\phi} B_{4,X} \right)
+2\dot{\phi}^2 \left( B_{4,\phi \phi}+\ddot{\phi} 
B_{4,X \phi} \right) \nonumber \\
& & +\frac{\ddot{\phi}^2}{\dot{\phi}^2} 
\left[ 2\dot{\phi}^2 B_{1,X} (1+2B_4) 
+2B_1 \left( 2\dot{\phi}^2B_{4,X} -1-2B_4 \right)
-3B_1^2 \left( 1+2B_4 \right)
\right]+4 \ddot{\phi} B_1 B_{4,\phi} \nonumber \\
& &+\frac{2}{\dot{\phi}} (1+2B_4) 
\left( B_{1,\phi} \dot{\phi} \ddot{\phi}
+B_1 \dddot{\phi} \right)\,, 
\label{Pphi}
\ea
\end{widetext}
where 
\ba
B_1 &\equiv& \frac{2XB_{4,X}}{1+2B_4}\,,
\label{B1}\\
P_m &\equiv& n_0 \rho_{m,n}-\rho_m\,,
\label{Pm}
\ea
with $X=\dot{\phi}^2/2$.
If $B_4$ depends on $\phi$ alone, both $A_4$ and $B_1$ 
vanish. In this case, the Lagrangian (\ref{Lag}) 
reduces to a sub-class of Horndeski theories.
Thus, the dimensionless variable $B_1$ characterizes the 
deviation from Horndeski theories.

The expressions of $\rho_{\phi}$ and $P_{\phi}$ 
derived above are valid even for DHOST theories 
with non-vanishing functions $A_3$ and $A_5$, 
in which case the function $B_1$ is given by 
$B_1=2X(B_{4,X}+A_3 X)/(1+2B_4)$~\cite{Cris18}. 
Since we are now considering the theories with $A_3=0$, 
$B_1$ is directly related to $B_4$ according to Eq.~(\ref{B1}).

On using the property 
$\dot{\rho}_m=\rho_{m,n} \dot{n}_0
+Q(\phi) \rho_m \dot{\phi}$ and 
the matter pressure (\ref{Pm}), 
the conservation (\ref{n0eq}) of total fluid number 
translates to
\ba
\dot{\rho}_{m}+3H \left( 1+w_{m}
\right) \rho_{m}=Q(\phi) \rho_m \dot{\phi}\,,
\label{back3}
\ea
where $w_m=P_m/\rho_m$.
Varying the total action ${\cal S}$ with respect to 
$\phi$, it follows that
\be 
\dot{\rho}_{\phi}+3H \left( 1+w_{\phi}
\right)\rho_{\phi}=-Q(\phi) \rho_m \dot{\phi}\,,
\label{back4}
\ee
where $w_{\phi}=P_{\phi}/\rho_{\phi}$. 
One can also derive Eq.~(\ref{back4}) by 
taking the time derivative of Eq.~(\ref{rhophi}) and 
using Eqs.~(\ref{Pphi}) and (\ref{back3}).

From Eq.~(\ref{back1}), the density parameters 
$\Omega_{\phi}=\rho_{\phi}/(3H^2)$ and 
$\Omega_{m}=\rho_{m}/(3H^2)$ obey
\be
\Omega_{\phi}+\Omega_{m}=1\,.
\label{Omepm}
\ee
{}From Eq.~(\ref{back2}) with Eq.~(\ref{back1}), 
we obtain
\be
h \equiv 
-\frac{\dot{H}}{H^2}=\frac{3}{2} 
\left( 1+w_{\rm eff} \right)\,,
\label{weff}
\ee
where $w_{\rm eff}$ is the effective equation 
of state defined by 
\be
w_{\rm eff}=w_{\phi}\Omega_{\phi}
+w_{m}\Omega_{m}\,.
\label{weff2}
\ee
We note that there are time derivatives $\dddot{\phi}$ in Eqs.~(\ref{back1})-(\ref{back2})
as well as $\ddddot{\phi}$ and $\ddot{H}$ in Eq.~(\ref{back4}).
As we will discuss in Sec.~\ref{fixedsec}, however, the background equations reduce to 
the dynamical system containing the time derivatives 
of $\phi$ and $a$ 
up to second order thanks to the degeneracy conditions. 

\section{Tracker and scaling solutions for $Q=0$}
\label{Sec:lag1}

We derive the Lagrangian $L$ allowing for the tracking solution satisfying
\be
\frac{\dot{\phi}}{H^p}=\alpha\,,
\label{dphiH}
\ee
where $p$ and $\alpha$ are constants. 
We focus on the case
\be
Q=0\,,
\ee
and impose the condition 
\be
w_{\phi}=
\frac{P_{\phi}}{\rho_{\phi}}={\rm constant}\,,
\ee
so that $P_{\phi}$ scales in the same way as $\rho_{\phi}$.

The tracker solution found for covariant Galileons~\cite{DT10,DT10b} 
corresponds to $p=-1$, with the field equation of state $w_{\phi}=-2$ 
during the matter era. 
The scaling solution found for cubic-order Horndeski 
theories~\cite{Frusciante:2018aew} corresponds to $p=1$, 
with $w_{\phi}=w_m$. 
Now, we are extending the analysis to a more general power $p$.

We take into account the canonical kinetic term $X$ in $G_2$ 
and search for the theories in which each term in $\rho_{\phi}$ 
and $P_{\phi}$ evolves in the same way as $X$, i.e., 
\be
\rho_{\phi} \propto P_{\phi} \propto \dot{\phi}^2
\propto H^{2p}\,.
\label{rhopro}
\ee
The terms associated with the couplings $G_2$ and $B_4$ 
in the Lagrangian (\ref{Lag2}) appear in the expressions 
of $\rho_{\phi}$ or $P_{\phi}$. 
Moreover, the $G_3$-dependent contributions to 
Eq.~(\ref{Lag2}) reduce to the term $-\dot{\phi}^2 
(G_{3,\phi}+\ddot{\phi}\,G_{3,X})$ in $P_{\phi}$ 
after the integration by parts. 
Then, the Lagrangian should follow the same time 
dependence as $\rho_{\phi}$ and $P_{\phi}$, i.e., 
\be
L \propto H^{2p}\,.
\label{Lpro}
\ee
In the following, we obtain the form of the Lagrangian 
allowing for the property (\ref{Lpro}). 
Since there are terms in $\rho_{\phi}$ and $P_{\phi}$ 
which are absent in $L$, we need to confirm whether 
each term in $\rho_{\phi}$ and $P_{\phi}$ obeys the 
property (\ref{rhopro}) after deriving the Lagrangian satisfying 
the condition (\ref{Lpro}).

The relation (\ref{Lpro}) translates to 
\be
\frac{\dot{L}}{HL}=-2ph\,,
\label{Lre}
\ee
where $h$ is defined by Eq.~(\ref{weff}).
In what follows, we consider the case in which 
$h$ is (nearly) constant. The constancy of $h$ exactly holds for
scaling solutions along which both $\Omega_{\phi}$ 
and $\Omega_{m}$ are constant.
For tracking solutions in which $\Omega_{\phi}$ varies 
in time, the quantity $h$ is approximately constant 
during the radiation- and matter-dominated epochs in which 
the contribution of the term $w_{\phi}\Omega_{\phi}$ 
to Eq.~(\ref{weff}) can be negligible. 
The constancy of $h$ also holds for the scalar-field dominated 
solution ($\Omega_{\phi}=1$).

In DHOST theories given by Eq.~(\ref{Lag}), the Lagrangian 
depends on $\phi$, $X$, $\square \phi$, $R$, and $Z$.
Then, we can write Eq.~(\ref{Lre}) in the form 
\ba
&&\frac{\partial L}{\partial \phi} 
\frac{\dot{\phi}}{H}+
\frac{\partial L}{\partial X} 
\frac{\dot{X}}{H}+
\frac{\partial L}{\partial \square \phi} 
\frac{\dot{(\square \phi)}}{H}
+\frac{\partial L}{\partial R}
\frac{\dot{R}}{H}
+\frac{\partial L}{\partial Z}
\frac{\dot{Z}}{H} \nonumber \\
&&=-2ph L\,.
\label{pareq}
\ea
On using the relation (\ref{dphiH}), 
the quantities associated with the time derivatives 
of $\phi$, $X$, $\square \phi=-\ddot{\phi}-3H \dot{\phi}$, 
$R=6(2H^2+\dot{H})$, and $Z=-\dot{\phi}^2 \ddot{\phi}^2$ 
can be expressed, respectively, as 
\ba
& &
\frac{\dot{\phi}}{H}=\frac{2ph}{\lambda} X^n\,,\\
& &
\frac{\dot{X}}{H}=-2ph X\,,
\label{dotX}\\
& &
\frac{\dot{(\square \phi)}}{H}
=-(p+1)h\square \phi\,,\\
& &
\frac{\dot{R}}{H}
=-2hR\,,\\
& &
\frac{\dot{Z}}{H}
=-2(2p+1)h Z\,, 
\ea
where 
\be
\lambda \equiv 2^{1-n}ph \alpha^{-1/p}\,,\qquad
n\equiv\frac{p-1}{2p}\,.
\ee
Substituting the Lagrangian (\ref{Lag}) into 
Eq.~(\ref{pareq}) and treating $A_4$ as 
an independent function from $B_4$, 
it follows that the couplings 
$G=G_2, G_3, B_4, A_4$ 
need to separately obey the partial differential equations: 
\be
XG_{,X}-\frac{1}{\lambda}X^n G_{,\phi}
-s G=0\,,
\label{par}
\ee
where $s$ is a constant given by 
\ba
s=
\begin{cases}
1 & \quad{\rm for }\quad G=G_2\,,\\
\dfrac{p-1}{2p} & \quad {\rm for }\quad G=G_3\,,\\
\dfrac{p-1}{p} & \quad {\rm for }\quad G=B_4\,,\\
-\dfrac{p+1}{p} & \quad {\rm for }\quad G=A_4\,.
\end{cases}
\label{sv}
\ea
In the following, we discuss the cases $p \neq 1$ and
$p=1$, separately.

\subsection{Tracker solutions: $p \neq 1$}
\label{Sec:tracker}

For $p \neq 1$ (i.e., $n \neq 0$), the general solution 
to Eq.~(\ref{par}) is given by 
\be
G(\phi,X)=X^s g({\cal Y})\,,
\label{GphiX}
\ee
where $g$ is an arbitrary function of 
\be
{\cal Y}=X^n+n \lambda \phi\,.
\label{Xcon}
\ee
Since we are now considering the case in which 
$h=-\dot{H}/H^2$ is approximately constant, the 
integration of this relation gives 
\be
H=\frac{1}{h(t-t_0)}\,,
\ee
where $t_0$ is a constant.
Then, we can integrate Eq.~(\ref{dphiH}) to give
\be
\phi=\phi_0+\frac{\alpha}{h^p (1-p)}(t-t_0)^{1-p}\,,
\ee
where $\phi_0$ is an integration constant. 
Since $X^n=2^{-n}\alpha^{2n}h^{1-p}(t-t_0)^{1-p}$, 
it follows that ${\cal Y}=n\lambda \phi_0={\rm constant}$.
Then, the function $g({\cal Y})$ does not vary in time 
along the tracker solution.

Let us study whether each term in $\rho_{\phi}$ and 
$P_{\phi}$ following from the Lagrangian (\ref{GphiX}) 
obeys the property (\ref{rhopro}). 
First of all, the quadratic Lagrangian is given by 
$G_2=Xg_2({\cal Y})$, where $g_2({\cal Y})$ is an arbitrary function 
of ${\cal Y}$. Since $g_2({\cal Y})$ does not vary in time along the 
tracker solution, the term $G_2$ in $\rho_{\phi}$ and 
$P_{\phi}$ evolves as $G_2 \propto X \propto H^{2p}$.
The contribution $\dot{\phi}^2 G_{2,X}$ to $\rho_{\phi}$ 
has the dependence $\dot{\phi}^2 G_{2,X}=\dot{\phi}^2(g_2+nX^n g_{2,{\cal Y}})$, 
so it satisfies the property (\ref{rhopro}) 
for $g_{2,{\cal Y}}=0$, i.e., $g_2({\cal Y})=c_2={\rm constant}$.
Hence the quadratic Lagrangian obeying the 
relation (\ref{rhopro}) is constrained to be 
\be
G_2=c_2 X\,.
\ee
The integrated solution to (\ref{par}) for
the cubic Lagrangian is given by 
$G_3=X^n g_3({\cal Y})$. 
In order to have the relation 
$\dot{\phi}^2 G_{3,\phi} \propto \dot{\phi}^2$, 
we require that 
$G_{3,\phi}=n \lambda X^n g_{3,{\cal Y}}$ does not 
change in time and hence $g_3({\cal Y})=c_3={\rm constant}$.
This restricts the Lagrangian to the form
\be
G_3=c_3 X^{(p-1)/(2p)}\,.
\label{G3con}
\ee
In this case, both the terms $3H \dot{\phi}^3 G_{3,X}$ and 
$-\dot{\phi}^2 \ddot{\phi} G_{3,X}$ are proportional 
to $\dot{\phi}^2$, so all the cubic-order contributions 
to $\rho_{\phi}$ and $P_{\phi}$ satisfy the relation 
(\ref{rhopro}). 
The cubic Galileon $G_3=c_3 X$ corresponds to $p=-1$, 
in which case the tracker solution characterized by 
$\dot{\phi}H={\rm constant}$ is present during the 
radiation- and matter-dominated epochs \cite{DT10,DT10b}.

The coupling $B_4$ following from the solution (\ref{GphiX}) 
is given by $B_4=X^{2n}b_4({\cal Y})$. 
The term $H \dot{\phi} B_{4,\phi}$ in $\rho_{\phi}$ and $P_{\phi}$ 
is in proportion to $H^{3p-1} b_{4,{\cal Y}}$. Since we are considering the 
case $p \neq 1$, this term is consistent with the dependence 
(\ref{rhopro}) only for $b_4({\cal Y})=c_4={\rm constant}$. 
Then we have 
\be
B_4=c_4 X^{(p-1)/p}\,,
\ee
so the function $B_1$ defined by Eq.~(\ref{B1}) yields 
\be
B_1=\frac{4n c_4 X^{2n}}{1+2c_4 X^{2n}}\,.
\ee

The field density $\rho_{\phi}$ and the pressure $P_{\phi}$ 
contain the terms like $H^2 B_4$ proportional to $H^{2p}$. 
On the other hand, there exists the term
\be
B_1^2 (1+2B_4) \frac{\ddot{\phi}^2}{\dot{\phi}^2}
=\frac{2^{4-n}n^2 c_4^2\,p^2 h^2 \alpha^{8n}}
{1+2c_4 X^{2n}}H^{4p-2}\,,
\label{B1B4}
\ee
which does not behave as $\propto H^{2p}$ for $p\neq 1$.
Under the condition $|2c_4 X^{2n}| \ll 1$, there are 
also contributions to $\rho_{\phi}$ and $P_{\phi}$ 
proportional to $c_4^3 H^{6p-4}$.
If we demand the {\it exact} tracking behavior along which 
all the terms in $\rho_{\phi}$ and $P_{\phi}$ 
have the dependence $\propto H^{2p}$, we have $c_4=0$
and hence
\be
B_4=0\,,\qquad B_1=0\,.
\ee
This property can be confirmed by substituting $B_4=c_4 X^{2n}$
into the degeneracy condition (\ref{A4B4}), i.e., 
\be
A_4=\frac{12n^2c_4^2}
{1+2c_4 X^{2n}}X^{-2/p}\,.
\ee
For $c_4 \neq 0$, this is at odds with the integrated solution 
$A_4=X^{-(p+1)/p}a_4({\cal Y})$.

Even in the case $c_4 \neq 0$, there exists an approximate 
tracker solution in the early cosmological epoch. 
Provided that $|2c_4 X^{2n}| \ll 1$, the leading-order terms to 
$\rho_{\phi}$ and $P_{\phi}$ for $p<1$ correspond to 
those proportional to $H^{2p}$, which arise from the 
couplings $G_2=c_2 X$, $G_3=c_3X^{(p-1)/(2p)}$ as well as 
$B_4=c_4X^{(p-1)/p}$. The existence of the coupling 
$B_4=c_4X^{(p-1)/p}$ gives rise to terms with the different 
power-law dependence of $H$. The next-to-leading 
contributions to $\rho_{\phi}$ and $P_{\phi}$
are in proportion to $H^{4p-2}$. 
Then, it follows that 
\ba
\rho_{\phi} &=& \alpha_1 H^{2p}
+c_4 \left( \alpha_2 H^{4p-2}+\cdots \right)\,,
\label{rhodeap}\\
P_{\phi} &=& \beta_1 H^{2p}
+c_4 \left( \beta_2 H^{4p-2}+\cdots \right)\,,
\label{Pdeap}
\ea
where $\alpha_{1,2}$ and $\beta_{1,2}$ are constants, and 
the abbreviation means the terms which 
are next order to $H^{4p-2}$.
Substituting Eqs.~(\ref{rhodeap})-(\ref{Pdeap}) and the time 
derivative $\dot{\rho}_{\phi}$ into the continuity equation 
$\dot{\rho}_{\phi}+3H(\rho_{\phi}+P_{\phi})=0$, we can solve 
it for $\beta_1$. 
On using this relation, the field equation of state 
$w_{\phi}=P_{\phi}/\rho_{\phi}$ yields 
\be
w_{\phi} \simeq -1+\frac{2}{3}ph
+c_4 \frac{2h(p-1) \alpha_2 H^{2(p-1)}}
{3\alpha_1+3c_4 \alpha_2 H^{2(p-1)}}\,,
\label{wphiana}
\ee
where we picked up the terms up to the order $H^{4p-2}$ 
in Eqs.~(\ref{rhodeap}) and (\ref{Pdeap}).

For $c_4=0$, we have $w_{\phi}=-1+2ph/3$. 
Indeed, the cubic Galileon corresponds to $p=-1$, in which 
case $w_{\phi}=-7/3$ during the radiation dominance ($h=2$) 
and $w_{\phi}=-2$ during the matter dominance 
($h=3/2$)~\cite{DT10,DT10b}.
For general values of $p$, we have $w_{\phi}=-1+p$ during 
the matter era. In this case, for $p$ closer to 0, 
the model can be compatible with 
the observational data associated with the background 
expansion history.

The non-vanishing coupling $B_4=c_4 X^{(p-1)/p}$ gives rise to 
the variation of $w_{\phi}$. 
For $p<1$, the terms in the parentheses of Eqs.~(\ref{rhodeap}) 
and (\ref{Pdeap}) evolve faster than $H^{2p}$, 
so they are suppressed relative to the former in the asymptotic 
past. In this limit, we recover the tracker equation of  
state $w_{\phi}=-1+2ph/3$. 
As long as the terms in the parentheses of Eqs.~(\ref{rhodeap}) 
and (\ref{Pdeap}) catch up with their first terms, 
$w_{\phi}$ starts to deviate from 
the tracker value $-1+2ph/3$.
Thus, in the presence of the coupling $B_4=c_4 X^{(p-1)/p}$, 
the tracking behavior can be approximately realized 
in the early cosmological epoch during which the terms proportional 
to $H^{4p-2}$ and $H^{6p-4}$ are subdominant to the 
$H^{2p}$ contributions to $\rho_{\phi}$ and $P_{\phi}$.

From Eq.~(\ref{wphiana}), we observe that, in the limit $p \to 1$,  
the field equation of state reduces to the tracker 
value $w_{\phi} \to -1+2h/3=w_{\rm eff}$ even in the presence of 
the DHOST term $B_4=c_4 X^{(p-1)/p}$. 
This limit corresponds to the scaling solution along which 
$w_{\phi}$ is equivalent to $w_m$ with constant $\Omega_{\phi}$. 
For $p=1$, the solution to Eq.~(\ref{par}) is different from 
Eq.~(\ref{GphiX}), so we discuss this case separately 
in the following.

\subsection{Scaling solutions: $p=1$}\label{Sec:scaling}

If $p=1$, then the solution to Eq.~(\ref{par}) 
is given by  
\be
G(\phi,X)=X^sg (Y)\,,
\label{Gsca}
\ee
where $s$ is given by Eq.~(\ref{sv}), and 
$g$ is an arbitrary function of 
\be
Y=X e^{\lambda \phi}\,,
\ee
and $\lambda$ is a constant. Each coefficient in the Lagrangian can be written in the form:
\ba
G_2(\phi,X) &=& Xg_2(Y)\,,\label{G2sca}\\
G_3(\phi,X) &=& g_3(Y)\,,\label{G3sca}\\
B_4(\phi,X) &=& b_4(Y)\,,\label{B4sca}\\
A_4(\phi,X) &=& X^{-2} a_4(Y)\,.\label{A4sca}
\ea
{}From the degeneracy condition (\ref{A4B4}),
the function $a_4(Y)$ is determined from $b_4(Y)$, as
\be
a_4(Y)=\frac{3Y^2 b_{4,Y}(Y)^2}{1+2b_4(Y)}\,.
\label{a4sca}
\ee
Unlike the case $p \neq 1$, the Lagrangian $A_4$ derived 
above is consistent with the other scaling Lagrangian $B_4$.
{}From Eq.~(\ref{B1}), the quantity $B_1$ is given by 
\be
B_1=\frac{2Yb_{4,Y}(Y)}{1+2b_4(Y)}\,,
\label{B1sca}
\ee
which depends on $Y$ alone.
For the solution satisfying the condition (\ref{dphiH}), 
the scalar field evolves as
\be
\phi=\phi_0+\alpha \ln a\,,
\ee
where $\phi_0$ is an integration constant.
Since $X \propto H^2$ and $e^{\lambda \phi} \propto 
a^{2h} \propto H^{-2}$, the quantity $Y$ 
remains constant. 
Then, the functions $G_3, B_4, B_1$ do not vary in time 
in the scaling regime.
On using the solutions (\ref{G2sca})-(\ref{B4sca}) and 
(\ref{B1sca}) in Eqs.~(\ref{rhophi}) and (\ref{Pphi}), 
we find that all the terms in $\rho_{\phi}$ and $P_{\phi}$
are in proportion to $H^2$.
Hence the background equations of motion obey the scaling 
property $\dot{\phi} \propto H$ for the Lagrangians 
(\ref{G2sca})-(\ref{A4sca}) with (\ref{a4sca}).

\section{Scaling Lagrangian for general 
field-dependent coupling $Q(\phi)$}
\label{Sec:lag2}

In Sec.~\ref{Sec:lag1}, we showed that the power $p=1$ 
is the special case in which all the terms of the background 
equations of motion scale in the same manner 
($\rho_{\phi} \propto P_{\phi} \propto H^2$) 
for $Q=0$. Now, we extend the analysis to the 
field-dependent coupling $Q(\phi)$ and derive the 
Lagrangian whose equations of motion obey the scaling 
relations $\rho_{\phi} \propto P_{\phi} \propto H^2$. 
Since the scaling solution satisfies the relation 
$\rho_{\phi}/\rho_m={\rm constant}$, both $\Omega_{\phi}$ 
and $\Omega_m$ are constant. 
Then, the effective equation of state $w_{\rm eff}$ 
and the quantity $h=-\dot{H}/H^2$ do not vary in time 
in the scaling regime.

\subsection{Derivation of the scaling Lagrangian}

The scaling relation $\rho_{\phi}/\rho_m={\rm constant}$ 
translates to $\dot{\rho}_{\phi}/\rho_{\phi}=\dot{\rho}_m/\rho_m$.
Then, from Eqs.~(\ref{back3}) and (\ref{back4}), we have
\be
\frac{\dot{\phi}}{H}=
\frac{2h}{\tilde{\lambda} Q(\phi)}\,,
\label{dotphi}
\ee
where
\be
\tilde{\lambda} \equiv \frac{2h}
{3\Omega_{\phi}(w_m-w_{\phi})}\,.
\ee
In the scaling regime, the quantity $\tilde{\lambda}$ 
is constant. {}From Eq.~(\ref{dotphi}), the field derivative has 
the dependence $\dot{\phi} \propto H/Q(\phi)$.

As we mentioned in Sec.~\ref{Sec:lag1}, the Lagrangian $L$ 
contains terms which are present in $\rho_{\phi}$ and $P_{\phi}$. 
We first derive the form of $L$ consistent with the condition 
$L \propto H^2$ and study whether some additional conditions 
are required to satisfy the scaling relations of each term 
in $\rho_{\phi}$ and $P_{\phi}$.
Then, the Lagrangian obeys
\be
\frac{\dot{L}}{HL}=-2h\,,
\label{Lsca}
\ee
which corresponds to $p=1$ in Eq.~(\ref{pareq}).
On using Eq.~(\ref{dotphi}), it follows that 
\ba
& &
\frac{\dot{X}}{H}=-2h
\left( 1+\frac{2Q_{,\phi}}{\tilde{\lambda} Q^2} 
\right)X\,,
\label{dotX}\\
& &
\frac{\dot{(\square \phi)}}{H}
=-2h \left( 1+{\cal F}_1 \right)\square \phi\,,\\
& &
\frac{\dot{R}}{H}=-2hR\,,\\
& &
\frac{\dot{Z}}{H}
=-2h \left( 1+{\cal F}_2\right)Z\,,
\ea
where 
\ba
\hspace{-0.5cm}
{\cal F}_1
&=&\frac{2(h-3)\tilde{\lambda}Q^2 Q_{,\phi}
-4h(Q Q_{,\phi \phi}-3Q_{,\phi}^2)}
{2\tilde{\lambda} Q^2 [(h-3)\tilde{\lambda} Q^2
+2hQ_{,\phi}]}\,,\\ 
\hspace{-0.5cm}
{\cal F}_2
&=&\frac{2Q^2(4\tilde{\lambda}Q_{,\phi}+\tilde{\lambda}^2 Q^2)
-4(QQ_{,\phi \phi}-4Q_{,\phi}^2)}{\tilde{\lambda} Q^2 (\tilde{\lambda}Q^2
+2Q_{,\phi})}\,.
\ea
Substituting the Lagrangian (\ref{Lag}) into Eq.~(\ref{pareq}), 
it follows that the couplings $G=G_2, G_3, B_4, A_4$ 
need to separately obey the partial differential equations: 
\be
\left( 1+\frac{2Q_{,\phi}}{\tilde{\lambda}Q^2} \right) 
XG_{,X}-\frac{1}{\tilde{\lambda}Q}G_{,\phi}
+f(\phi)G=0\,,
\label{pardif}
\ee
where 
\ba
f(\phi)=
\begin{cases}
-1 & \quad{\rm for }\quad G=G_2\,,\\
{\cal F}_1 & \quad {\rm for }\quad G=G_3\,,\\
0 & \quad {\rm for }\quad G=B_4\,,\\
{\cal F}_2 & \quad {\rm for }\quad G=A_4\,.\\
\end{cases}
\ea
The integrated solution to Eq.~(\ref{pardif}) 
is generally given by 
\be
G(\phi,X)=\tg (\tY)\,
e^{\tilde{\lambda} \int f(\phi)Q(\phi) d\phi}\,,
\label{Gso}
\ee
where $\tg$ is an arbitrary function of 
\be
\tY=Q^2(\phi) X e^{\tilde{\lambda}\psi}\,,
\ee
and $\psi$ is defined by 
\be
\psi=\int Q(\phi) d\phi\,.
\label{psi}
\ee
From Eq.~(\ref{Gso}), each coupling is 
restricted to be
\ba
G_2(\phi,X) &=& X\tg_2(\tY)Q^2(\phi)\,,
\label{G2}\\
G_3(\phi,X) &=& \tg_3(\tY) \frac{Q^3(\phi)}{q_1(\phi)}\,,
\label{G3}\\
B_4(\phi,X) &=& \tb_4(\tY)\,,\label{B4}\\
A_4 (\phi,X) &=& X^{-2}  \ta_4(\tY)
\frac{Q^4(\phi)}{q_2^2(\phi)}\,,\label{A4s}
\ea
where $\tg_2, \tg_3, \tb_4, \ta_4$ are arbitrary 
functions of $\tY$, and 
\ba
\hspace{-0.5cm}
q_1(\phi) &=&  Q^2(\phi)+
\frac{2h}{\tilde{\lambda}(h-3)}Q_{,\phi}(\phi)\,,\\
\hspace{-0.5cm}
q_2(\phi) &=&  Q^2(\phi)
+\frac{2}{\tilde{\lambda}}Q_{,\phi}(\phi)\,.
\ea
The Lagrangians $G_2$ and $G_3$ agree with those 
derived in Refs.~\cite{Amendola06} 
and~\cite{Frusciante:2018aew}, respectively.

The couplings $A_4$ and $B_4$ are related to each other 
according to the degeneracy condition (\ref{A4B4}). 
On using Eqs.~(\ref{B4}) and (\ref{A4s}), it follows 
that $Q_{,\phi}/Q^2={\rm constant}$.
This is integrated to give
\be
Q(\phi)=\frac{1}{\mu_1 \phi+\mu_2}\,,
\label{Qphi}
\ee
where $\mu_1$ and $\mu_2$ are constants. 
Thus, the degeneracy condition restricts the coupling 
to be of the form (\ref{Qphi}). 
In this case, both $q_1(\phi)$ and $q_2(\phi)$ are 
proportional to $Q^2(\phi)$. 
Absorbing the proportionality constant into the 
definitions of $\tilde{g}_3 (\tilde{Y})$ and 
$\tilde{a}_4 (\tilde{Y})$, the Lagrangian 
corresponding to the functions 
(\ref{G2})-(\ref{A4s}) yields
\ba\label{eq:scalingQphi}
L &= &
X\tg_2(\tY)Q^2(\phi)-\tg_3(\tY) Q(\phi)\square \phi 
\nonumber \\
& &+\tb_4(\tY) R+ X^{-2}\ta_4(\tY) Z\,,
\label{Lge}
\ea
where, from the degeneracy condition (\ref{A4B4}), 
the function $\ta_4(\tY)$ is constrained to be
\be
\tilde{a}_4 (\tilde{Y})=
\frac{3\tY \tb_{4,\tY}^2(\tY)^2}
{1+2\tb_4(\tY)}\,.
\label{ta4}
\ee

{}From Eq.~(\ref{B1}), we have  
\be
B_1(\phi,X)=\tb_1 (\tY) \equiv
\frac{2\tY \tb_{4,\tY}(\tY)}
{1+2\tb_4(\tY)}\,,
\ee
which depends on $\tY$ alone.
On using Eq.~(\ref{dotphi}), we find that the quantity 
$\psi$ defined by Eq.~(\ref{psi}) has the dependence 
$\psi=(2h/\tilde{\lambda}) \ln a+\psi_0$, 
where $\psi_0$ is a constant. 
Then, we have $e^{\tilde{\lambda}\psi} \propto a^{2h} \propto 
H^{-2}$ and hence $\tY \propto Q^2 \dot{\phi}^2 H^{-2}={\rm constant}$.
This means that the couplings $B_4$ and $B_1$ 
do not change in time in the scaling regime. 

Exploiting Eq.~(\ref{dotphi}) together with the property 
$Q_{,\phi}/Q^2=-\mu_1={\rm constant}$, 
it follows that 
\be
\ddot{\phi} \propto \frac{H^2}{Q(\phi)}\,,\qquad 
\dddot{\phi} \propto \frac{H^3}{Q(\phi)}\,.
\ee
Moreover, the $\phi$ and $X$ derivatives of 
$B_4=\tb_4(\tY)$ have the dependence
\ba
& &
B_{4,\phi} \propto Q(\phi)\,,\qquad
B_{4,X} \propto X^{-1}\,, \nonumber \\
& &
B_{4,\phi \phi} \propto Q^2(\phi)\,,\qquad 
B_{4,X \phi} \propto Q(\phi) X^{-1}\,,
\ea
where the function $B_1$ also satisfies similar relations. 
Then, all the terms appearing in $\rho_{\phi}$ and $P_{\phi}$ 
are in proportion to $H^2$. 
We have thus shown that the Lagrangian (\ref{Lge}) 
with the coupling (\ref{Qphi}) ensures the 
existence of scaling solutions. 

For the Lagrangian $G_2=X \tg_2(\tY)Q^2(\phi)$, there exist 
scaling solutions for any arbitrary coupling $Q(\phi)$. 
In theories containing the functions $G_3$ and $B_4$, 
however, the coupling is constrained 
to be of the form  (\ref{Qphi}). 
In cubic Horndeski theories the degeneracy conditions are absent, 
so the coupling $Q(\phi)$ does not seem to be constrained. 
Substituting Eq.~(\ref{G3}) into the $G_3$-dependent terms 
of $\rho_{\phi}$ and $P_{\phi}$, however, they are proportional 
to $H^2$ only for $Q_{,\phi}/Q^2={\rm constant}$. 
Hence we obtain the same coupling as Eq.~(\ref{Qphi})~\cite{Frusciante:2018aew}.

\subsection{Constant $Q$}

If $\mu_1=0$, then the matter coupling (\ref{Qphi}) 
is constant ($Q=1/\mu_2$).
Since $\psi=Q\phi$, the function (\ref{B4}) 
yields $B_4=\tilde{b}_4 (Q^2 Y)$, where 
\be
\lambda=\tilde{\lambda}Q=
\frac{2hQ}{3\Omega_{\phi}(w_m-w_{\phi})}\,,
\qquad Y=Xe^{\lambda \phi}\,.
\ee
We absorb the constant $Q^2$ into the new arbitrary 
function $b_4(Y)=\tb_4(Q^2 Y)$.
Applying the similar procedure to the other functions 
in Eq.~(\ref{Lge}), the existence of scaling solutions 
for non-vanishing constant $Q$ 
restricts the Lagrangian to be
\be
L=X g_2(Y)-g_3(Y)\square \phi 
+b_4(Y) R+ X^{-2}a_4(Y) Z\,,
\label{Lagcon}
\ee
where
\be
a_4(Y)=\frac{3Y^2 b_{4,Y}(Y)^2}{1+2b_4(Y)}\,,
\label{a4sca2}
\ee
with the function $B_1$ given by Eq.~(\ref{B1sca}).
The Lagrangian (\ref{Lagcon}) is of the same form as the one 
corresponding to the functions (\ref{G2sca})-(\ref{A4sca}) derived 
for $Q=0$. Thus, we have shown that the result (\ref{Lagcon}) 
is valid for both $Q=0$ and non-vanishing constant $Q$.

\subsection{Fixed points for constant $Q$}
\label{fixedsec}

We derive the fixed points for the dynamical system given by 
the Lagrangian (\ref{Lagcon}) with the application to 
dark energy in mind. 
The background Eqs.~(\ref{back1}) and (\ref{back2}) contain 
the time derivatives $\dot{H}$ and $\dddot{\phi}$, but they 
can be eliminated to give
\ba
& &
f_1(\dot{\phi},\phi)H^2 
+f_2(\ddot{\phi},\dot{\phi},\phi)H+
f_3(\ddot{\phi},\dot{\phi},\phi) \nonumber \\
& &
=\rho_m-3B_1 P_m\,,
\label{back3d}
\ea
where $f_1, f_2, f_3$ are functions of their arguments.
The branch with an expanding Universe corresponds to 
\be
H=\frac{\sqrt{f_2^2-4f_1 f_3+4f_1(\rho_m-3B_1 P_m)}-f_2}
{2f_1}\,.
\label{Hubble}
\ee
The quantity $f_2^2-4f_1 f_3$ does not possess the second 
derivative $\ddot{\phi}$, whereas $f_2/(2f_1)$ contains 
the term proportional to $\ddot{\phi}$.
Then, the Hubble parameter can be expressed in the form 
\be
H={\cal A}_1(\dot{\phi},\phi) \ddot{\phi}
+{\cal A}_2(\dot{\phi},\phi, \rho_m)\,,
\label{Hex}
\ee
where $P_m$ is related to $\rho_m$ according to 
$w_m=P_m/\rho_m={\rm constant}$.
Taking the time derivative of Eq.~(\ref{back3d}) and 
using the continuity Eq.~(\ref{back3}),
$\dot{H}$ and $\dddot{\phi}$ appear again.
However, they can be eliminated by 
using Eq.~(\ref{back1}). 
The resulting equation can be combined with Eq.~(\ref{back3d})
to solve for the second-order field 
derivative $\ddot{\phi}$ in the form 
\be
\ddot{\phi}={\cal B}_1 (\dot{\phi}, \phi, \rho_m)\,,
\label{ddotphi}
\ee
where we do not write the explicit form of ${\cal B}_1$ 
due to its complexity.
Substituting Eq.~(\ref{ddotphi}) into Eq.~(\ref{Hex}), 
it follows that the right hand side of Eq.~(\ref{Hex}) 
depends on $\dot{\phi}$, $\phi$, and $\rho_m$ alone.
Taking the time derivative of $H$ and using 
Eq.~(\ref{ddotphi}) again, we can express $\dot{H}$ 
in the form
\be
\dot{H}={\cal B}_2 (\dot{\phi}, \phi, \rho_m)\,.
\label{dotHf}
\ee
The above discussion shows that the dynamical system is 
kept up to second order in time derivatives 
for both $\phi$ and $a$.

To derive the fixed points of DHOST theories given by 
the Lagrangian (\ref{Lagcon}), it is convenient to introduce 
the following dimensionless variables:
\be
x \equiv \frac{\dot{\phi}}{\sqrt{6}H}\,,\qquad 
y \equiv \frac{e^{-\lambda \phi/2}}{\sqrt{3}H}\,,
\label{xydef}
\ee
where the quantity $Y$ 
can be expressed as $Y=x^2/y^2$.
Since both $x$ and $Y$ are constant along 
the scaling solution, $y$ does not 
vary in time either.
The variables $x$ and $y$ obey 
\ba
x' &=& x \left( \epsilon_{\phi}+h \right)\,,\label{dxeq}\\
y' &=& -y \left( \frac{\sqrt{6}}{2}\lambda x 
-h\right)\,,\label{dyeq}
\ea
where 
$\epsilon_{\phi}=\ddot{\phi}/(H \dot{\phi})$ and 
$h=-\dot{H}/H^2$, and a prime represents a derivative with respect to 
$N=\ln a$. The quantities $\epsilon_{\phi}$ and $h$ are 
known from Eqs.~(\ref{ddotphi}) and (\ref{dotHf}).

The fixed points of the dynamical system (\ref{dxeq})-(\ref{dyeq}) 
can be derived by setting $x'=0$ and $y'=0$.
The scaling solution obtained for constant $Q$ corresponds to  
\be
-\epsilon_{\phi}=h=\frac{\sqrt{6}}{2}\lambda x_c\,,
\label{eph}
\ee
where the subscript ``$c$'' represents the value on the 
critical point. Since the relations $\ddot{\phi}=-\sqrt{6}H^2h x_c$ 
and $\dddot{\phi}=2\sqrt{6}H^3 h^2 x_c$ hold on the fixed points, 
we substitute them into Eqs.~(\ref{back1})-(\ref{back2}) 
and solve them for $g_{2,Y}$ and $b_4$, respectively. 
On using Eq.~(\ref{ddotphi}), the fixed points satisfying 
the condition (\ref{eph}) obey
\be
\left[ 2(Q+\lambda) x_c-\sqrt{6} (1+w_m) \right]
\Omega_m=0\,.
\ee
There are the following two fixed points.

\begin{itemize}
\item (a) Scaling solution: $x_c=\dfrac{\sqrt{6} (1+w_m)}{2(Q+\lambda)}$\,.

This corresponds to the case in which $\Omega_{\phi}$ and $\Omega_m$ 
are non-vanishing constants. Along this solution, $w_{\phi}$ and 
$w_{\rm eff}$ are given, respectively, by 
\ba
w_{\phi} &=& w_m-\frac{Q(1+w_m)}{(1-\Omega_m)(Q+\lambda)}\,,\\
w_{\rm eff} &=& -\frac{Q-w_m \lambda}{Q+\lambda}\,.
\ea
For the vanishing coupling ($Q=0$), it follows that 
$w_{\phi}=w_{\rm eff}=w_m$. 
In the presence of the coupling $Q$, the scaling solution
can lead to the cosmic acceleration for $w_{\rm eff}<-1/3$, 
but we need  $|Q|$ to be larger than the order $|\lambda|$ for achieving this purpose.

\item (b) Scalar-field dominated point: $\Omega_m=0$\,.

There is another fixed point satisfying $\Omega_{\phi}=1$.
In this case, we have 
\be
w_{\rm eff}=w_{\phi}=-1+\frac{\sqrt{6}\lambda x_c}{3}\,,
\ee
where $x_c$ is known for given functions $g_2(Y)$, $g_3(Y)$, 
and $b_4(Y)$. 
If $\lambda x_c<\sqrt{6}/3$, then the point (b) can be 
used for the late-time cosmic acceleration.
\end{itemize}

For the dynamical system (\ref{dxeq})-(\ref{dyeq}), 
there exist other kinetic-type fixed points satisfying
\be
y_c=0\,,
\ee
under which $Y_c=x_c^2/y_c^2 \to \infty$.
The functions $g_2(Y), g_3(Y), b_4(Y)$ consistent with 
the background equations of motion are given by 
\ba
g_2(Y) &=& \sum_{n \geq 0} c_n Y^{-n}\,,\label{g2con}\\
g_3(Y) &=& \sum_{n \geq 1} d_n Y^{-n}\,,\\
b_4(Y) &=& \sum_{n \geq 1} e_n Y^{-n}\,,\label{g4con}
\ea
where $c_n, d_n, e_n$ are constants and $n$ is an integer.  
In $g_3(Y)$, we do not include a constant $d_0$ since 
it is just a total derivative.
The term $\tilde{d}_1\ln Y$ can be taken into account 
in $g_3(Y)$ as in Refs.~\cite{Alb,Frusciante:2018aew}. 
Here, we do not 
do so since we are interested in the effect of the function 
$b_4(Y)$ on the fixed points. 
A constant $e_0$ is not included in $b_4(Y)$ 
by reflecting the fact that this is merely 
a shift of the reduced Planck mass.  
We substitute Eqs.~(\ref{g2con})-(\ref{g4con}) and 
their $Y$ derivatives into Eqs.~(\ref{back1})-(\ref{back2}) 
and solve them for $\Omega_m$ and $h$. 
Plugging these relations into Eq.~(\ref{ddotphi}), 
we find that there are the following two fixed points.

\begin{itemize}
\item (c) $\phi$MDE\,.

This is characterized by 
\be
x_c=\frac{\sqrt{6}}{3c_0 (w_m-1)}Q\,,
\ee
with 
\ba
w_{\phi} &=&1\,,\\
w_{\rm eff} &=& 
w_m-\frac{2Q^2}{3c_0(w_m-1)},\\
\Omega_{\phi} &=& 
\frac{2Q^2}{3c_0 (w_m-1)^2}\,.
\ea
The constant $e_n$ ($n \geq 1$) in $b_4(Y)$ does not 
modify the values of $w_{\phi}$, $w_{\rm eff}$, and 
$\Omega_{\phi}$ of the standard $\phi$MDE \cite{Amen00}. 
For $w_m=0$, we have 
$w_{\rm eff}=\Omega_{\phi}=2Q^2/(3c_0)$.
Provided that $|Q| \ll 1$, the 
$\phi$MDE can replace the standard matter era.

\item (d) Purely kinetic point.

There exists another kinetic point satisfying 
\be
x_c=\pm \sqrt{\frac{1}{c_0}}\,,
\ee
with 
\be
w_{\phi}=1\,,\quad w_{\rm eff}=1\,,\quad
\Omega_{\phi}=1\,.
\ee
This can be used for neither radiation/matter eras 
nor the cosmic acceleration. 
\end{itemize}

In cubic-order Horndeski theories, it was shown in 
Ref.~\cite{Frusciante:2018aew} that there exist
viable dark energy models with the $\phi$MDE 
followed by the fixed point (b). 
In the presence of the coupling $b_4(Y)$ of 
the form (\ref{g4con}), it is of interest to study 
in detail how the cosmological dynamics and the evolution 
of perturbations are subject to change 
compared to cubic-order Horndeski theories.

\section{Conclusions}
\label{consec}

In this paper, we considered quadratic-order DHOST theories satisfying degeneracy conditions to avoid the Ostrogradsky instability, the constraint on the speed of gravitational waves, and the bound on the decay of 
gravitational waves to dark energy perturbations. 
The Lagrangian of this class is given by Eq.~(\ref{Lag}), where $A_4$ 
is related to $B_4$ according to Eq.~(\ref{A4B4}).
We derived the most general Lagrangians that are able to reproduce separately 
tracking and scaling behaviors under the condition that 
$h=-\dot{H}/H^2$ is approximately constant.
In the absence of coupling $Q$ between the scalar field and matter, 
we obtained the Lagrangian of tracking solutions satisfying the 
conditions $L \propto H^{2p}$ and $\dot{\phi} \propto H^p$.
In particular, the scaling behavior corresponds to the choice $p=1$. 

In Sec.~\ref{Sec:tracker}, we showed that, for $Q=0$, 
the exact tracker solution exists up to the cubic-order Horndeski 
Lagrangian with the functions 
$G_2=c_2X$ and $G_3=c_3X^{(p-1)/(2p)}$. 
We verified  that these contributions to the background equations 
obey the relations $\rho_\phi \propto P_\phi \propto H^{2p}$. 
In the presence of the DHOST Lagrangian, we found that the function 
$B_4=c_4X^{(p-1)/p}$ leads to the approximate tracker solution 
at early times when the terms proportional to $H^{2p}$ (with $p<1$)  
are the dominant contributions to $\rho_{\phi}$ and $P_{\phi}$.
At late times, the other terms in $\rho_{\phi}$ and $P_{\phi}$, which 
grow faster than $H^{2p}$, give rise to a variation in the field 
equation of state $w_{\phi}$.  For $p=1$, we found that the exact 
scaling solution can be realized by the DHOST Lagrangian given by 
Eqs.~(\ref{G2sca})-(\ref{A4sca}) with (\ref{a4sca}).

In Sec.~\ref{Sec:lag2}, we  extended the analysis of scaling solutions
to the case of a field-dependent coupling $Q(\phi)$. 
The most general Lagrangian with scaling solutions is of the form 
(\ref{Lge}), with $\tilde{a}_4 (\tilde{Y})$ related to 
$\tilde{b}_4 (\tilde{Y})$ according to Eq.~(\ref{ta4}).
We showed that the degeneracy condition (\ref{A4B4}) fixes the form 
of the coupling to be $Q(\phi)=1/(\mu_1\phi+\mu_2)$, 
including the constant $Q$ as a special case. 
Indeed, we verified that all the terms in $\rho_{\phi}$ and $P_{\phi}$ 
are in proportion to $H^2$.  
The coupling $Q(\phi)$ can be arbitrary for the quadratic Lagrangian 
$L=X \tilde{g}_2 (\tilde{Y})Q^2(\phi)$ alone, but the existence of 
cubic and quartic Lagrangians restricts the coupling to be
of the above form to satisfy the scaling property of each term 
in $\rho_{\phi}$ and $P_{\phi}$ (as shown in 
Ref.~\cite{Frusciante:2018aew} for the cubic Lagrangian).

For a non-vanishing constant $Q$, the Lagrangian with scaling 
solutions reduces to the form (\ref{Lagcon}) with (\ref{a4sca2}), which matches 
with the result found for $Q=0$ in Sec.~\ref{Sec:scaling}. 
In Sec.~\ref{fixedsec}, we derived the fixed points of the dynamical system described by this Lagrangian. In particular, we obtained four fixed points: 
(a) a scaling critical point, 
(b) a scalar-field dominated point, 
(c) a $\phi$MDE point, and 
(d) a purely kinetic critical point. 
The points (c) and (d) arise for the models given by  
the functions (\ref{g2con})-(\ref{g4con}).
The point (a) is unlikely to be responsible for the late-time cosmic 
acceleration with $w_{\rm eff}$ close to $-1$ because one would need a large value for the coupling $|Q|$, while the observations of temperature anisotropies 
in cosmic microwave background place the upper bound 
$|Q|< {\cal O}(0.1)$ \cite{Ade:2015rim}. 
On the other hand, the other scaling point (c) can replace
the standard matter era. Moreover, the point (b) can be used 
for driving the cosmic acceleration.

It would be of interest to apply the Lagrangians with tracking and 
scaling solutions to the construction of concrete dark energy models. 
In particular, one can investigate whether there exists a viable cosmology 
allowing for the $\phi$MDE point (c) followed by the accelerated point (b) 
without ghost and Laplacian instabilities.
In such a case, one can explore the differences with the 
dark energy model in cubic-order Horndeski theories
where a viable cosmological sequence 
exists~\cite{Frusciante:2018aew}.  
The analysis of cosmological perturbations is also important 
to compare those models with the observations associated with 
the cosmic growth history.

\acknowledgments
ST thanks warm hospitalities during his stay in University of Portsmouth at which a part of this work was done. 
The research of NF is supported by Funda\c{c}\~{a}o para a  Ci\^{e}ncia e a Tecnologia (FCT) through national funds  (UID/FIS/04434/2013), by FEDER through COMPETE2020  (POCI-01-0145-FEDER-007672) and by FCT project ``DarkRipple -- Spacetime ripples in the dark gravitational Universe" with ref.~number PTDC/FIS-OUT/29048/2017. 
RK is supported by the Grant-in-Aid for Young 
Scientists B of the JSPS No.\,17K14297.   
KK is supported by the UK STFC grant ST/N000668/1 and the European Research Council under the European Union's Horizon 2020 programme (grant agreement No.\,646702 ``CosTesGrav").
ST is supported by the Grant-in-Aid for Scientific Research Fund of the 
JSPS No.~16K05359 and MEXT KAKENHI Grant-in-Aid for 
Scientific Research on Innovative Areas ``Cosmic Acceleration'' (No.\,15H05890). 
DV acknowledges financial support from the FCT Project 
No.\, UID/FIS/00099/2013.


\end{document}